\documentclass[a4paper,fleqn,usenatbib]{mnras}

\usepackage{newtxtext,newtxmath}

\usepackage[T1]{fontenc}
\usepackage{ae,aecompl}

\usepackage{graphicx}	
\usepackage{amsmath}	
\usepackage{amssymb}	

\title[Upper-Limit Lensing] {`Upper-Limit Lensing': Constraining galaxy stellar masses with singly-imaged background sources}
\author[Russell J. Smith et al.]{
Russell J. Smith\thanks{E-mail: russell.smith@durham.ac.uk}
John R. Lucey
and William P. Collier
\\
Centre for Extragalactic Astronomy, University of Durham, Durham DH1 3LE, United Kingdom\\
}

\date{Accepted 2018 August 23. Received 2018 August 22; in original form 2018 July 13}

\pubyear{2018}

\begin{document}
\label{firstpage}
\pagerange{\pageref{firstpage}--\pageref{lastpage}}
\maketitle

\begin{abstract}

Strong gravitational lensing can provide accurate measurements of the stellar mass-to-light ratio $\Upsilon$ in low-redshift ($z$\,$\la$\,0.05) early-type galaxies, and hence probe
for possible variations in the stellar initial mass function (IMF).
However, true multiple imaging lens systems are rare, hindering the construction of large nearby lens samples.
Here, we present a method to derive upper limits on $\Upsilon$ in galaxies with single
close-projected background sources, where no counter-image is detected, down to some relative flux limit.
We present a proof-of-principle application to three galaxies with integral field observations from different instruments. 
In our first case study, only a weak constraint on $\Upsilon$ is obtained. In the second, the absence of a detectable counter-image excludes stellar masses higher than expected for a Salpeter IMF. 
In the third system, the current observations do not yield a useful limit, but our analysis indicates that deeper observations should reveal a counter-image if the stellar mass is any larger  
than expected for a Milky Way IMF. 
We discuss how our method can help enlarge the current samples of low-$z$ galaxies with lensing constraints,
both by adding upper limits on $\Upsilon$ and by guiding follow-up of promising single-image systems in search of fainter counter-images.
\end{abstract}

\section{Introduction}

Strong gravitational lensing can be exploited on galaxy scales to provide robust measurements of total projected mass within a characteristic aperture, with  uncertainties of only a few per cent 
\citetext{e.g. \citealt{2010ARA&A..48...87T}}. If the relative contributions of dark matter and stars can be estimated, e.g. using additional information from stellar dynamics, then the 
stellar mass-to-light ratio $\Upsilon$ can be determined, which in turn constrains the stellar initial mass function (IMF). Applications of this method to lenses at redshift $z$\,=\,0.2 
provided some of the first evidence for `heavy' IMFs in  elliptical galaxies \citep{2010ApJ...709.1195T}, with a factor of $\sim$2 mass excess, compared to Milky-Way-like IMFs, for the most
massive objects (velocity dispersions $\sigma$\,$\ga$\,300\,km\,s$^{-1}$).

In recent work, we have developed the use of lenses at lower redshifts ($z$\,\la\,0.05) as an especially robust probe of the IMF, minimising the uncertainties associated with the dark-matter component
\citep*{2013MNRAS.434.1964S}. 
This advantage arises because the critical surface density $\Sigma_{\rm cr}$ (in M$_\odot$\,pc$^{-2}$) 
scales inversely with the lens distance (assuming distant sources), so that low-redshift lensing 
can occur only in the high density central parts of galaxies, where stars dominate the total mass. In this case, the stellar mass-to-light ratio  $\Upsilon$ can be estimated from a `pure' lensing analysis, 
without using any dynamical information. From four systems studied in detail to date, with $\sigma$\,$\approx$\,300\,km\,s$^{-1}$, \cite{2018MNRAS.478.1595C} concluded that $\Upsilon$ 
is on average only 9$\pm$8 per cent heavier than expected from a \cite{2001MNRAS.322..231K} IMF. This result differs strikingly from the distant lensing analysis, and is also in tension with 
results from pure stellar dynamics \citep{2013MNRAS.432.1862C} and spectroscopic limits on low-mass stars \citep{2012ApJ...760...71C}.

As the small sample size in \cite{2018MNRAS.478.1595C} suggests, identifying nearby lenses is challenging. The number of potential lenses 
is limited by the small volume available at low redshift, and the intrinsic rarity of high-mass galaxies. The number of nearby galaxies with {\it bright} lensed background objects, 
recognisable as such from broad-band imaging,
is prohibitively small. Instead, we have developed a strategy using integral field unit (IFU) spectroscopy to detect multiply-imaged line-emission from faint sources behind selected nearby galaxies. 
Our method was first applied in the SNELLS (SINFONI Nearby Elliptical Lens Locator Survey) programme \citep*{2015MNRAS.449.3441S}, and later adapted to optical data from 
MUSE (Multi-Unit Spectroscopic Explorer)  \citep{2018MNRAS.478.1595C}, leading to the sample of four multiple-imaging lenses described above, as well as a number of 
close-projected background emitters without any identifiable lensed counter-image.

The new generation of multi-IFU galaxy surveys offers a promising new route to augmenting the nearby lens sample.
The ongoing  SAMI (Sydney-AAO Multi-object Integral field spectrograph) survey \citep{2015MNRAS.447.2857B} and SDSS-IV MaNGA 
(Mapping Nearby Galaxies at Apache Point) \citep{2015ApJ...798....7B} are observing thousands of low-redshift 
galaxies, including hundreds which are massive enough potentially to act as strong lenses. 
\cite{2018MNRAS.477..195T} recently identified 40 background emission-line sources behind MaNGA target galaxies.
In most of these systems, only a single image of the background source is detected in the MaNGA data-cube\footnote{Talbot et al. report multiple imaging for nine background emitters
based on pseudo-narrow-band images from the pipeline datacubes. We have independently analysed the data for these systems, using the method of \cite{2017MNRAS.464L..46S}, which was 
developed to avoid artifacts near galaxy centres caused by the cube reconstruction process. With this analysis, only two of the Talbot et al. systems show possible evidence for multiple imaging: 
SDSSJ170124.01+372258.09  \citep[as reported in][]{2017MNRAS.464L..46S}
and SDSSJ143607.49+494313.22, both at $z_{\rm lens}$\,$\sim$\,0.12.}, but future
follow-up observations could detect counter-images to establish more of these systems as low-redshift lenses.

In this paper, motivated by the difficulty of finding multiple-imaging lenses at low redshift, we consider whether useful lensing information can be inferred in the 
cases noted above where only a {\it single} (and unresolved) image of a background source is detected in close projection. As a crude argument (neglecting detection thresholds, external shear, ellipticity, etc), 
consider a background emitter observed at a projected separation $r_{\rm em}$ from the lens candidate. Lensed counter-images should be formed for any image 
projected within twice the Einstein radius, $R_{\rm Ein}$. Hence if {\it no} counter-image is present, 0.5\,$r_{\rm em}$ must be larger than the 
Einstein radius, and therefore enclose a mean surface density which is smaller than the critical value, i.e. $\langle\Sigma\rangle_{0.5 r_{\rm em}}$\,$<$\,$\Sigma_{\rm cr}$. 
The critical surface density $\Sigma_{\rm cr}$ depends only on the foreground and background redshifts, which are known (in the case of IFU lens searches). 
Hence if the lensing mass is dominated by 
stars, so that a mass-follows-light model is applicable, this calculation yields an upper limit to the stellar mass-to-light ratio $\Upsilon$. 

An alternative, pixel-based, method for exploiting singly-imaged systems was described by \cite{2015ApJ...803...71S}, and applied to mainly more distant 
galaxies from the SLACS (Sloan Lens Advanced Camera for Surveys) project. As we discuss later in this paper, while Shu et al.'s goals
are similar to ours, their implementation is quite different, and not well-suited to the systems discussed here, where the lensed images are unresolved.

The remainder of this paper expands on the simplified argument above, taking into account some of the complicating factors already noted.
In Section~\ref{sec:meth} we present a more general treatment of a circular `toy model', illustrating the effect of a finite detection threshold, 
such that only sufficiently bright counter-images are detectable, and the impact of the unknown external contributions to the lensing deflections, modelled as a quadrupole shear.
In  Section~\ref{sec:applic}, we illustrate the method with application to three observed galaxies (accounting now also for the lens ellipticity), drawn from different data sources, 
which exemplify the range of results that can be obtained. 
We show that in favourable cases, useful upper limits to $\Upsilon$ can be obtained, which in principle provide further information on the IMF mass excess factor in elliptical galaxies. 
In other cases, our method provides a framework to guide future observations, with the goal of 
either establishing new multiply-imaged systems or else placing tighter upper limits on $\Upsilon$ from single images.
Section~\ref{sec:disc} summarizes the results and discusses the prospects for using `upper-limit lensing' to derive statistical results for larger samples from current and future surveys. 

Where necessary, we adopt a cosmology with parameters ($h$, $\Omega_{\rm M}$, $\Omega_{\Lambda}$)\,=\,(0.696, 0.286, 0.714) \citep{2014ApJ...794..135B}.
If we had instead adopted the \cite{2016A&A...594A..13P} parameters, with $h$\,=\,0.678, all determinations of $\Upsilon$  would be reduced by 2.5 percent.

\begin{figure}
\includegraphics[width=82mm,angle=270]{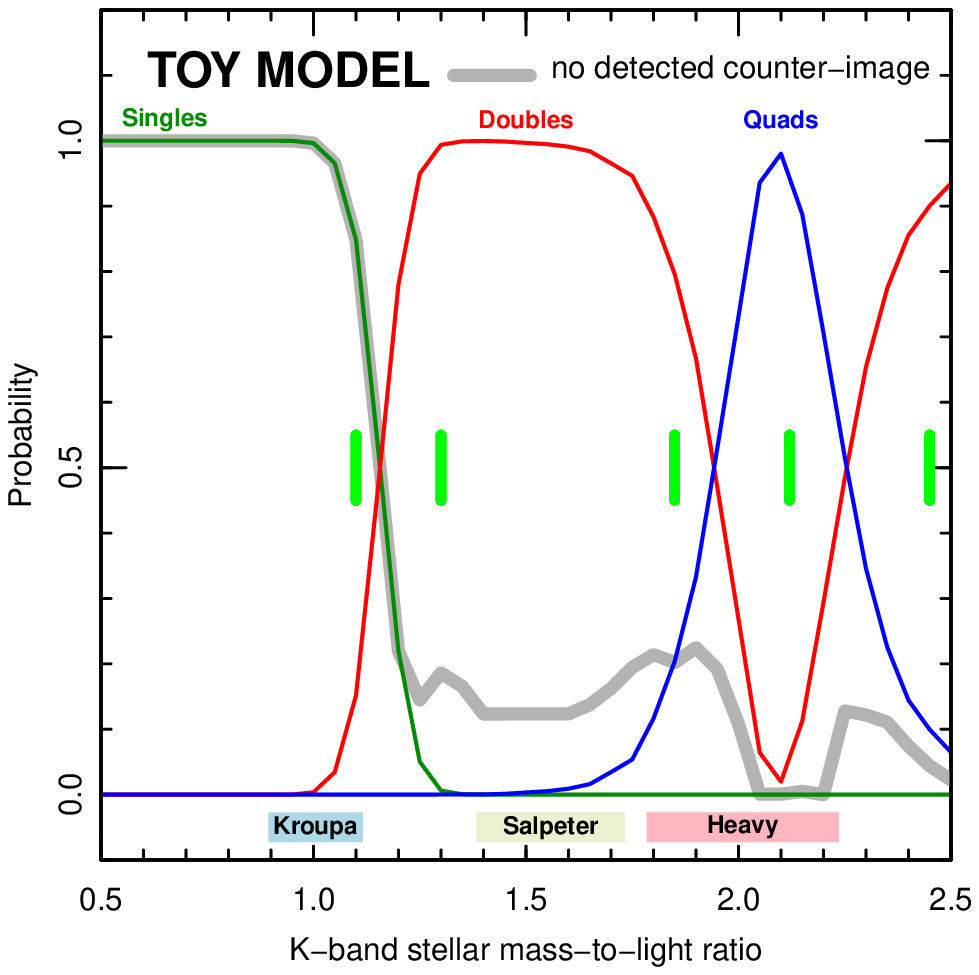}
\vskip -1mm
\caption{Upper-limit lensing constraints on the mass-to-light ratio, for the toy-model case in Section~\ref{sec:meth}. Here, the mass profile is a circular $R^{1/4}$-law profile, and we assume one
image of the background galaxy (the `test image') has been detected at a projected separation half of the effective radius.
Green, red and blue lines show the fraction of {\it intrinsically} single-, double- and quadruple-image configurations, for fixed observed location of the `test image',
as a function of the stellar mass-to-light ratio $\Upsilon$. 
The heavy grey line shows the probability of {\it not} observing any lensed counter-image, to a flux limit of one quarter that of the test image, i.e. $f_{\rm lim}$\,=\,0.25. The difference 
between green and grey lines is attributable to multiple image systems (mainly doubles) with counter-images falling below this limit. The green vertical bars show the values of 
$\Upsilon$ selected in Figure~\ref{fig:configs}, corresponding to different regimes in typical multiplicity. The horizontal scaling is arbitrary for this synthetic example, but set to be 
similar to the observed cases shown in subsequent figures. Below, we indicate the range in $\Upsilon$ spanned by old-metal-rich populations with a 
Milky-Way-like IMF, a Salpeter IMF ($\times$1.55 higher in $\Upsilon$),
and a `heavy' IMF ($\times$2 higher).}
\label{fig:unlens_circ}
\end{figure}

\begin{figure*}
\includegraphics[width=170mm,angle=270]{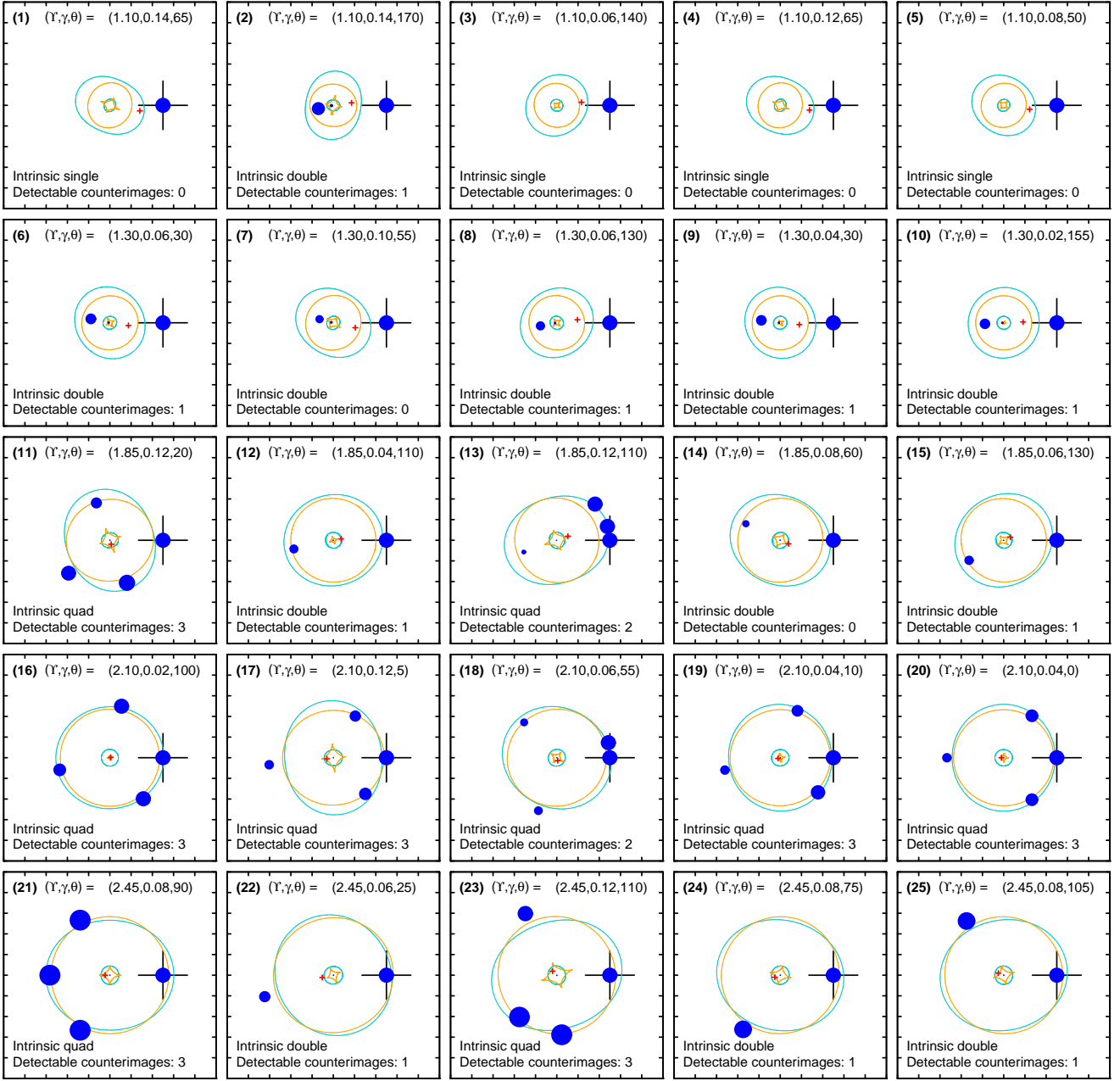}
\caption{Example image configurations for the circular toy-model case. The fixed `test image' position (representing the detected image of the background galaxy)
is at half the effective radius of the mass distribution, and indicated by the blue circle with cross hairs. Lensed counter-images are shown by other blue points, with size 
proportional to their
magnification, relative to the test image. The orange and cyan lines show lensing caustic and critical curves, respectively; the red cross indicates the undeflected source position.
Each row shows five cases for a given value of $\Upsilon$ (these values are indicated in Figure~\ref{fig:unlens_circ}).
The shear amplitude $\gamma$ and orientation $\theta$ are randomly drawn, to illustrate the scatter introduced by these unknown parameters. 
In each panel we note the {\it intrinsic} multiplicity (ignoring `central', i.e. Fermat-maximum, images), and also the 
number of {\it detectable} counter-images, defined as having $>$25\% the flux of the test image.}
\label{fig:configs}
\end{figure*}

\begin{table*}
\caption{
Relevant parameters of the systems analysed in Section~\ref{sec:applic}. 
The velocity dispersion $\sigma_{\rm 6dF/SDSS}$ is from  \protect\cite{2014MNRAS.443.1231C} or \protect\cite{2009ApJS..182..543A}.
}
\label{tab:snells}
\begin{tabular}{lcccl}
\hline
Short ID 				 & SNL-4  & PGC007748 & J0728+4005 & \\
\hline
2MASS ID (2MASX...)				& J04431291--1542101 & J02021739-0107405 & J07281702+4005025 & \\
IFU data source & SNELLS & MUSE & MaNGA \\
\hline
$z_{\rm lens}$ 		 & 0.037 & 0.043 & 0.050  \\
$z_{\rm src}$ 		 & 1.38 & 0.830 & 0.954 & \\ 
$\sigma_{\rm 6dF/SDSS}$ 	& 304$\pm$16 & 254$\pm$6 & 268$\pm$6 & km\,s$^{-1}$ \\
\hline
$R_{\rm eff}$ & 		3.8  & 11.6  & 5.4 & effective radius (arcsec)\\
PA & 	     --52.8 &  +72.8 & +69.6 & deg E of N\\
$e$  & 	    	0.38	& 0.20  & 0.21 & ellipticity 1--$b/a$\\
\hline
$L_{\rm ap}$ & 	 17.6 & 18.9 & 20.6 &  $K$-band; $R_{\rm ap}$\,=\,5$\arcsec$ ($10^{10}$\,$L_\odot$)\\
\hline
$D_A$  & 152.6  & 175.9 &  201.4 & ang. size distance of lens (Mpc) \\ 
$\Sigma_{\rm cr}$  & 11.34 & 10.09 &  8.778 & crit. surf. density ($10^3$\,$M_\odot$\,pc$^{-2}$) \\ 
\hline
$r_{\rm em}$  & 4.25 & 3.64  & 3.16 & emitter separation (arcsec)\\
PA$_{\rm em}$ & --36.0 & --77.6  & --53.2 & emitter angle (deg E of N) \\
$f_{\rm lim}$ &  0.4 & 0.08  & 0.4 & relative flux limit for counter-image \\
\hline
\end{tabular}
\end{table*}

\section{Method}\label{sec:meth}

In this section, we present a brief summary of how upper-limit lensing can be used to extract information about the stellar mass-to-light ratios in early-type galaxies.
We describe the method as implemented for the real galaxies in Section~\ref{sec:applic}, and illustrate it by application to a simple circular toy model.

Our key assumption is that the lensing mass is dominated by the stellar component of the foreground galaxy, 
and that this can be adequately described using a constant stellar mass-to-light ratio $\Upsilon$. 
While dark matter is evidently not negligible in many lensing situations 
(e.g. cluster lenses, or distant lenses with Einstein rings far larger than the effective radius), it should be a reasonable assumption
for low-redshift galaxies where $R_{\rm Ein}$\,$\la$\,0.5\,$R_{\rm eff}$, as in the cases presented in Section~\ref{sec:applic}.
For example, \cite{2015MNRAS.449.3441S} estimated contributions of $\la$20 per cent from cosmological simulations.
A constant {\it stellar} mass-to-light ratio has been assumed in most dynamical and lensing studies, 
although some recent works suggest a trend of increasing $\Upsilon$ towards galaxy centres, possibly related to IMF gradients 
\citep{2018MNRAS.476..133O,2018arXiv180101883S}.

For the present analysis we represent the stellar mass component with an elliptical $R^{1/4}$-law profile in the lensing model, but a more flexible profile could be adopted.
Dark matter (DM) is assumed to make only a small contribution to the lensing deflection within the radius probed by the emitter. 
To the extent that this is true, the difference in distribution of the DM, compared to that of the stars, is unlikely to affect the configuration significantly.
In this case, any DM contribution simply inflates the derived (upper) limit on $\Upsilon$.  An explicit DM halo, e.g. with a \cite*{1996ApJ...462..563N} profile, could be included 
in the calculation in principle.

The lensing configuration will be affected by the external mass distribution, which can be represented to first order by a quadrupole
`shear' term, parametrized by an amplitude, $\gamma$, and angle, $\theta$. 
For a given observed position and lens mass, the background galaxy may be singly imaged for some shear amplitudes and orientations, but multiply imaged for others.
Since  $\gamma$ and $\theta$ are unknown, the shear introduces a probabilistic element into the calculation, and the analysis must sample and marginalise over the likely range of values.

With the above assumptions, for given parameters of the foreground galaxy (redshift, effective radius, ellipticity, position angle, luminosity within some calibration aperture)
and of the background source (redshift and position of the observed emitter, relative to the lens), we can generate a large set of lens models, exploring 
the ($\Upsilon$, $\gamma$, $\theta$) parameter space.
We use the {\sc gravlens} code of \cite{2001astro.ph..2340K} to solve the lens equation for each model in this grid, all conditioned on the position of one observed image (the `test image') of
the background object. For each  ($\Upsilon$, $\gamma$, $\theta$), we compute the position of the background galaxy in the source plane, and locate any
corresponding counter-images in the image plane. 

It is important to distinguish between {\it intrinsically} singly-imaged systems, and those which {\it appear} single because any counter-images are fainter than the observational
flux limit. We define this threshold, $f_{\rm lim}$, relative to the test image, and apply it to the relative 
image magnifications predicted by the lens model. This is valid if the background source is small compared to variations in the magnification map, but becomes unreliable close
to the caustics, so we impose a cap of 20 on the absolute magnifications to suppress unrealistic flux ratios. 
Additionally, we include the size and orientation of the instrument field-of-view
so that any `outer' counter-images lost from the field are properly accounted for. 

The remainder of the calculation simply summarizes the image positions and fluxes from the model grid.
At each ($\Upsilon$, $\gamma$) pair, we determine the fraction of $\theta$ values yielding no {\it detectable} counter-images, $F_{\rm N}(\Upsilon, \gamma)$. 
Then, by marginalising over $\gamma$, we compute the probability of observing no counter image down to the imposed relative flux limit:
$U(\Upsilon)$\,=\,$\int F_{\rm N}(\Upsilon, \gamma) P(\gamma)\,$d$\gamma$.
Here $P(\gamma)$ is the probability distribution for the shear amplitude, which can 
be written as a 2D-Maxwellian distribution, i.e. $P(\gamma)$\,d$\gamma$\,$\propto$\,$\gamma\,e^{-\gamma^2/(2s^2)}$, 
where $s$ is the rms for each cartesian component of the shear vector. 

As a concrete example, Figure~\ref{fig:unlens_circ} shows the results derived for a circular toy model, where we posit a test image at half the 
effective radius, 0.5\,$R_{\rm eff}$, of the $R^{-1/4}$-law surface density profile. 
This represents the position of the observed background emitter in the image plane. 
The normalisation of the lensing mass is set by the free parameter, $\Upsilon$, and by a calibrating aperture luminosity; 
this luminosity is arbitrary for our toy model example, but chosen to yield results for $\Upsilon$ which are similar to those of the real galaxies described in the following section. 
For this example, we assume that no counter-image is detected down to a flux of $f_{\rm lim}$\,=\,0.25 relative to the test image. Again, this is similar to the values for the observed galaxies.
The grid of lensing models explored spans ranges $\Upsilon$\,=\,0.5--3.0 (step 0.05), $\gamma$=0.00--0.20 (step 0.02) and $\theta$\,=\,0--180$^\circ$ (step 5$^\circ$).
In marginalising over the shear amplitude, we assume $s$\,=\,0.05\, i.e. an rms of 5\,per cent in each component, which is
similar to the average derived from mass-follows-light + shear models of Grade A lenses in \cite{2008ApJ...682..964B}.
Some example lensing configurations from the model grid are shown in Figure~\ref{fig:configs}.

As expected, the likely intrinsic multiplicity of the system depends strongly on $\Upsilon$, with several regimes being identifiable. At low $\Upsilon$ ($<$1.15), intrinsically-single
images predominate, while double-image configurations\footnote{Strictly, because our lensing model does not diverge at the origin, these configurations have a third, highly demagnified, central image.
We formally include these `Fermat-maximum' images in the calculation, but they do not affect the results as they are rarely if ever above the flux limit, and never the {\it only} observable counter-image.}
are more likely at higher $\Upsilon$. Another transition occurs where quadruply-imaged systems become more likely than doubles 
(at $\Upsilon$\,=\,1.95). Doubles again become dominant at $\Upsilon$\,$>$\,2.25. The $\Upsilon$ ranges spanned by these regimes will in general depend on a combination of 
mass normalisation and projected separation of the background galaxy, while the sharpness of the transitions between them depends on the assumed shear rms.

In the intrinsically double configurations with 1$<$$\Upsilon$$<$2, the test image is the outer, generally brighter, of the pair.
The counter-image is closer to the lens galaxy centre and is usually fainter, so that it may fall below the relative flux threshold (e.g. Panel 7 of Figure~\ref{fig:configs}).
Hence in practice, $U(\Upsilon)$ falls off less rapidly than the fraction of true singles, the difference being attributable to faint 
counter-images\footnote{If a dark matter halo is present, with an inner density profile {\it shallower} than that of the stars \citep*[e.g.][]{1996ApJ...462..563N}, then these
inner counter-images have slightly {\it higher} relative flux; hence $U(\Upsilon)$ is conservatively over-estimated by neglecting the DM.}.
Quadruply-imaged systems generally have less disparate flux ratios, so $U$ falls quickly to zero in the quad-dominated regime\footnote{
We note however that marginal `fold' quads present a problem for our point-image-based analysis: in these configurations  the brightest of three counter-images  is 
formed very close to the test image, and the two may be blended and indistinguishable in the observations (e.g. Panel 18 of Figure~\ref{fig:configs}). A future refinement of the method could account for the image PSF, but 
the identification of such cases as lenses will also depend on the intrinsic source structure, so this is only a partial solution.}.
In our example case, where the mass profile is circular, and the shear rms is small, quads predominate in only a narrow range of $\Upsilon$; 
more elliptical and/or sheared models generate quads more efficiently.
In the intrinsically double configurations with $\Upsilon$$>$2, the test emitter is the inner image of the source, which is usually the fainter of the pair;
hence in these systems the counter-image is unlikely to fall below the detection limit (e.g. Panel 25 of Figure~\ref{fig:configs}).

If the results in Figure~\ref{fig:unlens_circ} were obtained for a real galaxy, we would conclude  
that observing no counter-image in this system would be somewhat unlikely ($<$20\,per cent)
for mass-to-light ratios larger than $\Upsilon$\,$\approx$\,1.3. Deeper `observations' would help to strengthen this statement, by ruling out the 
1.3$<$$\Upsilon$$<$2.0 models with fainter counter-images.

\section{Example applications}\label{sec:applic}

\begin{figure*}
\includegraphics[width=140mm,angle=270]{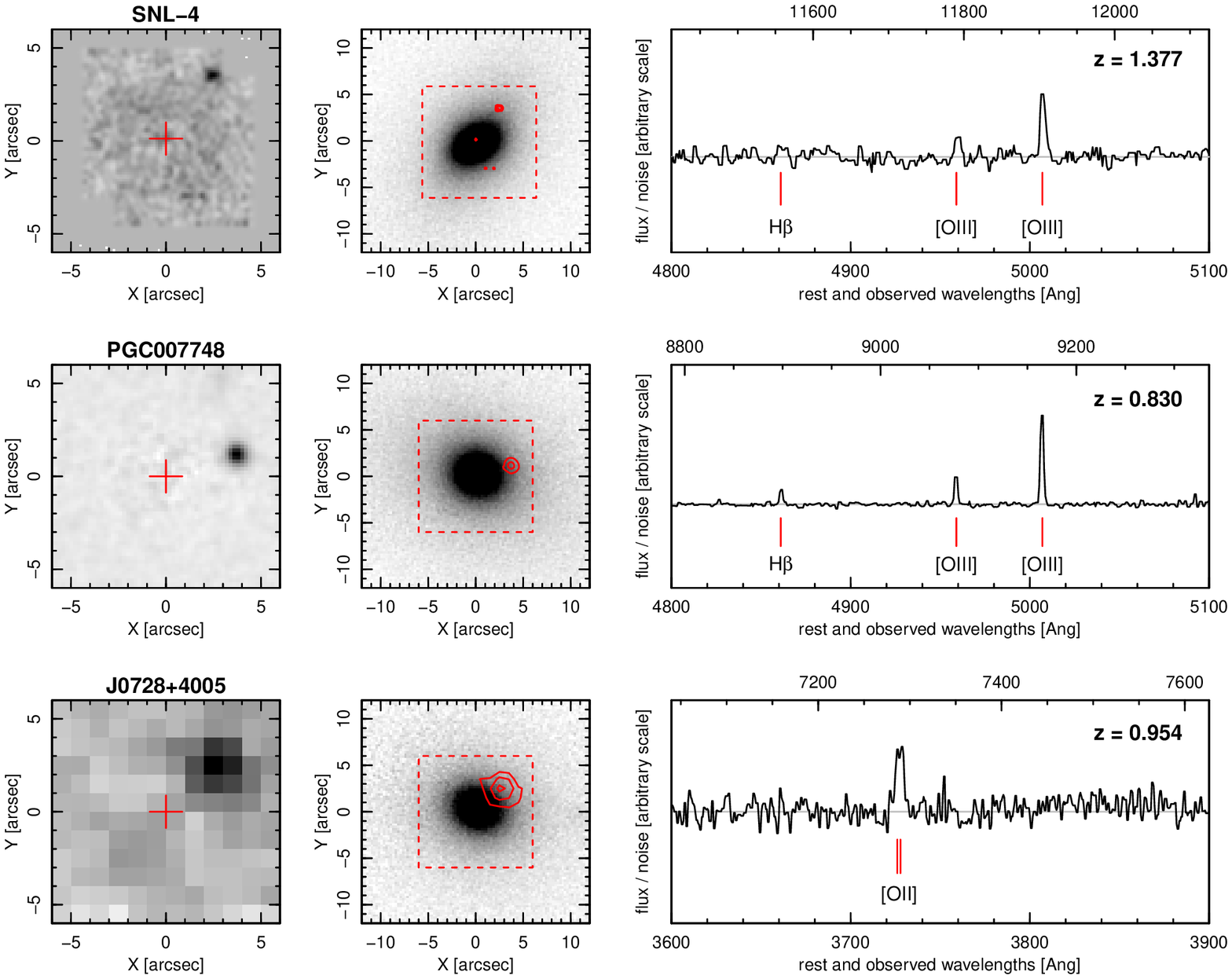}
\vskip -1mm
\caption{IFU observations of the three single-image systems analysed in this paper. For  SNL-4, we show SINFONI data from the SNELLS project. 
For P007748, the data are from archival MUSE observations. For J0728+4005, we use MaNGA data re-processed according to the methods of \protect\cite{2017MNRAS.464L..46S}.
In each row, the left-hand panel shows a narrow-band image extracted around the brightest line from the background emitter.
The second panel shows the PS1 $y$-band continuum image, with the emission line map overlaid as contours (the dashed outline indicates the box size in the left-hand panel).
The right-hand panel shows the residual spectrum extracted at the position of the line emitter and showing the brightest lines in each case.
}
\label{fig:sinfofig_extra_plus}
\end{figure*}

In this section we present a proof-of-principle implementation of our technique to three observed cases of single close-projected emitters behind massive early-type galaxies at $z$\,\la\,0.05.
Although the systems are drawn from three different IFU datasets (SNELLS, MUSE and MaNGA), we exploit uniform imaging survey data to keep the analysis as consistent as possible.

For applying the upper-limit lensing method, as currently implemented, the necessary inputs are: 
\begin{enumerate}
\item The redshifts of the foreground and background galaxy (for the distances entering $\Sigma_{\rm cr}$ in the lensing calculations),  and the position of the observed background emitter relative to the 
foreground galaxy. These parameters are derived from the IFU data. 
\item The effective radius, position angle, ellipticity and flux normalisation (defining the lensing potential, up to an unknown factor of $\Upsilon$).  These are obtained from fits to 
broad-band images. The flux normalisation is defined in $K$-band from 2MASS \citep{2006AJ....131.1163S}, to minimise sensitivity to age effects when comparing to the expected values for candidate IMFs. 
We use an aperture of 5\,arcsec radius, and include a (small) correction for the PSF. 
For the effective radius, position angle, and ellipticity, we use {\sc galfit} \citep{2010AJ....139.2097P} $R^{1/4}$-law fits to
$y$-band images from the Pan-STARRS (PS1) survey  \citep{2016arXiv161205560C}.  This profile is a good fit to the galaxies analysed here. 
\item An upper limit on the flux of any counter-image, relative to the observed image. This is obtained by adding
fake sources to the residual emission-line image derived from the IFU data and visually estimating the minimum flux at which an inner counter-image can be securely 
identified.
\item The definition of the model grid and assumed shear rms. These are set as in Section~\ref{sec:meth}, except where specified below.
\end{enumerate}

The relevant observational parameters for the three galaxies  are summarized in Table~\ref{tab:snells}.

Because the normalisation of the lensing model is applied through the $K$-band luminosity, all values of $\Upsilon$ refer to this band.
The limits on $\Upsilon$ can be interpreted relative to the expectations for different choices of the IMF. 
Ideally, this would take into account the metallicity and star-formation history of 
each galaxy, as determined from high signal-to-noise spectroscopy \citep[e.g.][]{2017ApJ...845..157N},
but such data are not available for all of the objects discussed here. Instead we simply compare to a plausible range in the 
expected `reference' stellar mass-to-light ratio $\Upsilon_{\rm ref}$, assuming that the stars are old and metal rich, as typical in massive ellipticals. 
Specifically, from the \cite{2005MNRAS.362..799M} models, populations with metallicity 1--2\,$Z_\odot$ and ages 10--13\,Gyr (formation epoch $z$\,$\ga$\,2) 
have $K$-band ratios 0.89\,$<$\,$\Upsilon_{\rm ref}$\,$<$\,1.12, for a \cite{2001MNRAS.322..231K} IMF.
The limits on $\Upsilon$ can then be posed in terms of the mass-excess factor $\alpha$\,=\,$\Upsilon/\Upsilon_{\rm ref}$, such that 
a Milky-Way-like IMF has $\alpha$\,=\,1.0, an unbroken \cite{1955ApJ...121..161S} IMF has $\alpha$\,=\,1.55, and we define a `heavy' IMF as $\alpha$\,=\,2.0.

\subsection{Case 1: SNL-4 from SNELLS}\label{sec:snl4}

The first case we consider is a previously unreported single-image system from the SNELLS programme.
In \cite{2015MNRAS.449.3441S}, we identified three multiple-image systems, and one close single-image background galaxy. The latter is unsuitable for our current analysis as it has a 
strong dust lane.  In subsequent campaigns, we observed 26 further candidate lens galaxies, using essentially the same observing strategy.
As before, targets were drawn from the 6dFGSv \citep{2014MNRAS.443.1231C} and SDSS \citep{2009ApJS..182..543A} surveys, 
selecting for high velocity dispersion $\sigma$\,$\ga$\,300\,km\,s$^{-1}$ and low redshift ($z$\,$<$\,0.065). 
Cluster members and central galaxies of rich groups were excluded, to minimise DM halo contributions.
No new multiply-imaged emitters were discovered from these observations, but one further 
close-projected background galaxy was identified, with unambiguous  $z$\,=\,1.38 [O\,{\sc iii}] emission detected behind the $z$\,=\,0.037 target 2MASX\,J04431291--1542101, hereafter SNL-4.

\begin{figure}
\includegraphics[width=82mm,angle=270]{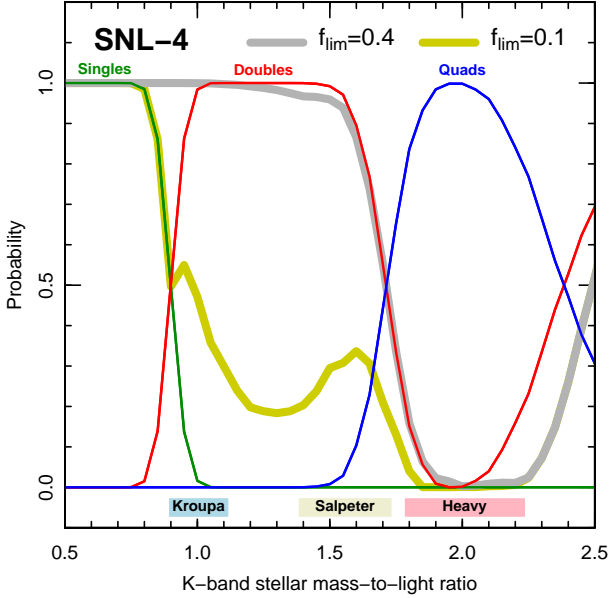} 
\vskip -1mm
\caption{Upper-limit lensing constraints on the $K$-band stellar mass-to-light ratio $\Upsilon$ in SNL-4. The heavy grey line shows the probability of {\it not} observing a
lensed counter-image to the emission-line source discovered in SNELLS, as a function of $\Upsilon$, for the relative flux limit of the existing data, $f_{\rm lim}$\,=\,0.4.
The yellow curve shows a hypothetical deeper observation with $f_{\rm lim}$\,=\,0.1.
The calculations account for the limited field-of-view, which causes the rise at large $\Upsilon$. Other annotations are as in Figure~\ref{fig:unlens_circ}.
The SNELLS observation only excludes very large values, $\Upsilon$\,$\ga$\,1.85, and even a significant increase in depth would not improve this constraint. 
The position of the background source along the major axis of SNL-4  leads to demagnified counter-images which hinder derivation of tight limits on $\Upsilon$ with our method.
}
\label{fig:unlens_snl34}
\end{figure}

The SINFONI data for SNL-4 are shown in the first row of Figure~\ref{fig:sinfofig_extra_plus}. Adding fake sources (with the appropriate PSF) to the narrow-band 
emission-line image, we estimate a relative flux limit of $f_{\rm lim}$\,=\,0.4. (Note that this estimate must be made in the image domain, and is hence larger
than a naive estimation from the significance of the line in the spectrum.)
The SNELLS data only cover two offset 8$\times$8\,arcsec$^2$ fields, so some models in the  ($\Upsilon$, $\gamma$, $\theta$) 
grid produce unobserved counter-images falling outside the field-of-view.

The results for SNL-4 are shown in Figure~\ref{fig:unlens_snl34}.  The {\it intrinsic} multiplicity curves are similar to those in the toy model case, except that
the quad-dominated regime spans a wider range in $\Upsilon$. This difference is attributable to the high ellipticity of the foreground galaxy, 
which enlarges the inner caustic enclosing quadruply-imaged sources. However, the {\it apparent}-singles curve $U(\Upsilon)$, which accounts for the detectability of the counter-image, is
quite different than for the toy model: for SNL-4, only the quadruple systems would be identified as lenses at the SNELLS depth. The counter-images in double systems would be too
faint to detect at any $\Upsilon$. This is partly due to the faintness of the test image itself, and correspondingly high value of $f_{\rm lim}$, and partly due to the system geometry:
the high ellipticity of the foreground galaxy, and the alignment of the source along its major axis, tends to increase the flux ratio between the two images, producing a 
relatively demagnified counter-image. 
The limit we infer on the mass-to-light ratio for SNL-4 is $\Upsilon$\,$<$\,1.85, defined by the point where the probability of observing no counter-image
falls below 0.1. The limited SNELLS field-of-view prevents detection of the widest doubles at large $\Upsilon$, hence $U(\Upsilon)$ rises above 0.1 again for $\Upsilon$\,$>$\,2.3, 
though such high values might be considered implausible {\it a priori}.

Hence we conclude that the absence of counter-image in SNL-4 is fully compatible with either  a Milky-Way-like or a Salpeter IMF. 
A correction for a $\sim$20 per cent dark-matter contribution inside the region studied would not qualitatively change this statement.
The results for this galaxy are limited largely by the difficulty of recovering faint counter-images in doubly-imaged configurations. 
Much deeper observations would be necessary to derive useful limits on the IMF
for this galaxy. (Figure~\ref{fig:unlens_snl34} shows a comparison case for $f_{\rm lim}$\,=\,0.1, 
which would still not be sensitive enough to detect many of the faint doubles.)

\subsection{Case 2: PGC007748 from MUSE}\label{sec:muse}

\begin{figure}
\includegraphics[width=82mm,angle=270]{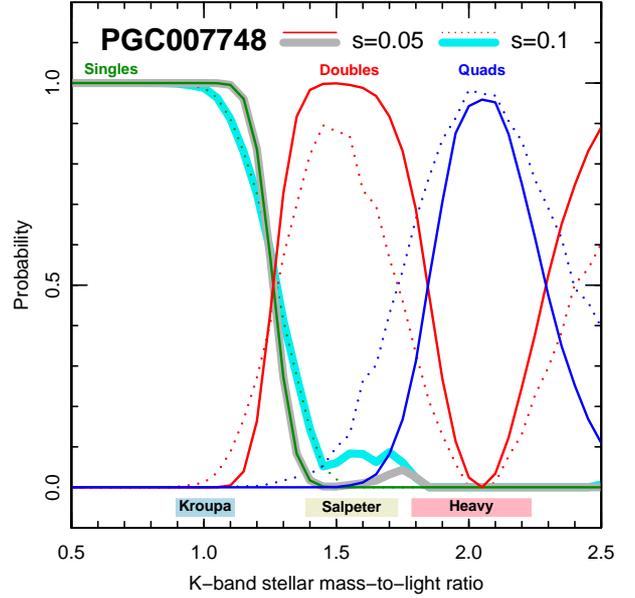} 
\vskip -1mm
\caption{Upper-limit lensing results for PGC007748, derived from MUSE data. The thick grey curve shows the probabilities derived for 5\% rms shear in each direction.
Since this is the central galaxy of a cluster (Abell 295), it may be subject to larger shear than the other cases considered in this paper; hence for comparison, we also 
show the equivalent results for a calculation where the assumed rms is doubled (dotted lines and thick cyan curve). In contrast to the SNL-4 case, the existing data provide a useful 
upper-limit constraint on $\Upsilon$: an IMF heavier than Salpeter is disfavoured in this galaxy by the absence of a detectable counter-image.}
\label{fig:unlens_p7}
\end{figure}

The second system is a close-projected emitter discovered in our systematic search for nearby lenses in public archival MUSE observations  \citep[][and in preparation]{2018MNRAS.478.1595C}. 
The foreground galaxy is PGC007748 (2MASX\,J02021739--0107405), which is the central galaxy of Abell 295, 
a poor cluster with X-ray mass $M_{200}$\,=\,6$\times$10$^{13}$\,M$_\odot$ \citep{2011A&A...534A.109P} and velocity dispersion $\sigma_{\rm cl}$\,=\,$359^{+52}_{-32}$\,km\,s$^{-1}$ 
\citep{1996ApJ...473..670F}.
This target was observed as part of the MUSE Most Massive Galaxies project (094.B-0592, PI: E. Emsellem), with  a total exposure time of 1.6\,hr and seeing $\sim$0.8\,arcsec FWHM.
Our processing follows the scheme described in \cite{2018MNRAS.478.1595C}, starting from the Phase 3 MUSE-DEEP datacube released by ESO.

Two apparently distinct emission-line sources were identified at a common redshift of $z$\,=\,0.830, at distances of 3.6\,arcsec and 5.5\,arcsec
from the galaxy centre. The inner feature is not seen in {\it Hubble Space Telescope} continuum imaging (500\,sec {\it F814W} WFPC2) from Laine et al. (2003), while the outer one is barely visible
there.
The two emission sources have very different spectra, with the inner one bright in  [O\,{\sc iii}], while the outer 
feature has strong [O\,{\sc ii}] but is much fainter in [O\,{\sc iii}] (seen at top of frame in Figure~\ref{fig:sinfofig_extra_plus}). 
Considering this difference, as well as the spatial configuration (both sources on the same side
of the target galaxy), we interpret the system as two separate singly-imaged objects at the same redshift (the line centres differ by $<$\,50\,km\,s$^{-1}$). 
Here, we use only the inner image to derive stellar mass constraints for PGC007748. 

In calculating the lensing model grid, we adopt a relative flux limit of $f_{\rm lim}$\,=\,0.08, estimated by adding fake sources to the emission-line image.
All counter-images are formed within the large field-of-view of MUSE (1\,arcmin). Given that this object may reside in a more 
massive halo, we explore two choices for the external shear rms: the default case, with $s$\,=\,0.05,
and a high-shear case, with $s$\,=\,0.10. As expected, the latter assumption leads to a greater blurring between the single, double and quad regimes. 

The results are shown in Figure~\ref{fig:unlens_p7}. The mass-to-light ratio constraint is much stronger than for the case of SNL-4, due to 
the bright background source and to the more circular mass distribution which favours relatively brighter counter-images.
For $s$\,=\,0.05, the probability of not detecting a counter-image falls below 0.1 at $\Upsilon$\,$\approx$\,1.4;
the constraint is only marginally weaker in the high-shear calculation.
Unlike the SNL-4 case, there is no rise in $U(\Upsilon)$ at large $\Upsilon$, due to the large field of view.
Comparing to the expected values for old stellar populations, this result disfavours a Salpeter IMF (or other IMF with similarly high mass-to-light ratio).
Correcting for DM contributions would lead to an even smaller maximum $\Upsilon$.
We note that deeper observations would not significantly alter the results for this system,
since the existing data can already detect the predicted counter-images for almost all of the intrinsically-multiple models in the grid.

\subsection{Case 3: J0728+4005 from MaNGA}\label{sec:mang}

Our final application is to one of the single-image systems from the MaNGA survey, reported by \cite{2018MNRAS.477..195T}.
Among this sample, J0728+4005 (2MASX\,J07281702+4005025) is the most suitable candidate for our method, because the background emitter is projected 
at very small separation (3.2\,arcsec), and the velocity dispersion of the foreground galaxy is relatively large ($\sigma$\,=\,268\,km\,s$^{-1}$). 
J0728+4005 does not appear to be part of any substantial group or cluster.

To establish the relative flux limit, we constructed a narrow-band image centred on the [O\,{\sc ii}] line, using the method described in \citep{2017MNRAS.464L..46S}. The key 
difference compared to the approach of Talbot et al. is that we remove the galaxy continuum signal from each fibre individually {\it prior} to reconstructing a residual data-cube, which produces
a much cleaner residual image, especially close to the centre of the foreground galaxy.
Adding fake sources with the MaNGA PSF (2\,arcsec FWHM), we estimate a relative flux limit for counter-image detection of $f_{\rm lim}$\,=\,0.4. 
The MaNGA fibre bundle used for this galaxy has a field-of-view diameter $\sim$20\,arcsec, which is sufficient to cover all counter-images generated in the lensing models.

\begin{figure}
\includegraphics[width=83mm,angle=270]{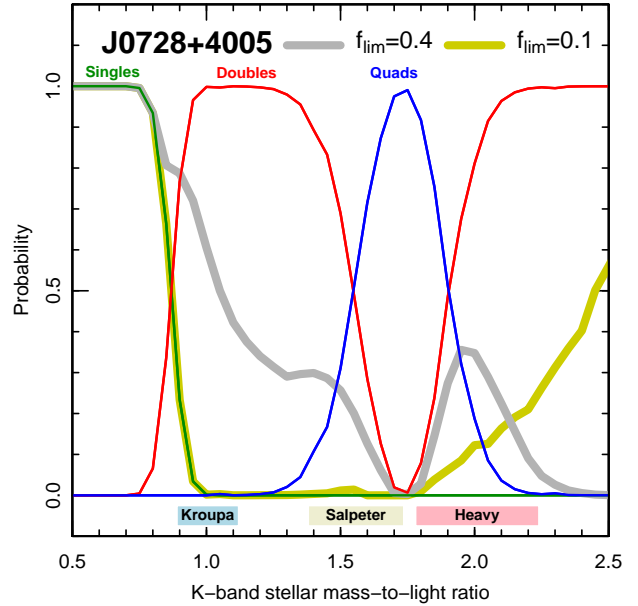} 
\vskip -1mm
\caption{Upper-limit lensing results for J0728+4005. The thick grey curve assumes the MaNGA field-of-view and relative detection threshold. 
The yellow curve is for a putative follow-up observation with the Gemini GMOS IFU, comprising three 5$\times$7\,arcsec$^2$ fields (see Figure~\ref{fig:pred}), and relative threshold of 0.1.
Although MaNGA does not yield a useful limit on $\Upsilon$, deeper observations of this system should reveal a counter-image, if the IMF in this galaxy is any heavier than that of the Milky Way.}
\label{fig:unlens_m}
\end{figure}

The results for J0728 are shown in Figure~\ref{fig:unlens_m}. The absence of a detectable counter-image in the MaNGA data places only weak constraints on $\Upsilon$;
only very high masses ($\Upsilon$\,>\,2.2) and a narrow interval in the quad-dominated regime ($\Upsilon$\,$\approx$\,1.7) are excluded. Hence little can be concluded 
about the IMF in this galaxy from the MaNGA data alone.

On the other hand, our analysis shows that J0728+4005 is an inherently powerful system deserving further study, due to the high luminosity of the foreground galaxy, and the
small projected separation of the background emitter. Moreover, this target does not suffer the combination of high ellipticity and alignment of the background source with the
lens major axis, which causes de-magnification of any counter-images in SNL-4.
Hence in Figure~\ref{fig:unlens_m}, we also show $U(\Upsilon)$ calculated for a possible future observation with $f_{\rm lim}$\,=\,0.1, which could be undertaken
for example with the Gemini GMOS IFU. This sensitivity gain is a realistic goal, given the increase in telescope size and achievable angular resolution compared to MaNGA.

With the assumptions of the deeper observation, essentially all doubles with inner counter-images should be detectable, so $U(\Upsilon)$ falls below 0.1 at $\Upsilon$\,$\approx$\,0.9.
This test shows that deeper observations should recover a counter-image to the background source for any model where the IMF is heavier than that of the Milky Way, adding a new 
low-redshift lens to the small sample currently in hand. Conversely, failure to detect the counter-image would be clear evidence for a `lightweight' (i.e. Milky-Way-like) IMF in this galaxy, 
even assuming negligible  DM contributions.

\section{Discussion and summary}\label{sec:disc}

We have described a new method to derive limits on the stellar mass-to-light ratios in galaxies with single
close-projected background sources, and presented a proof-of-principle application to  three galaxies from different IFU data sources. 

Our approach is most suitable for application to IFU observations of massive field galaxies, because in this case:
{\it (a)} the redshifts of both the foreground and background galaxy are known, and hence so is the critical density for lensing, $\Sigma_{\rm cr}$;
{\it (b)} the lensing convergence $\kappa$\,=\,$\Sigma/\Sigma_{\rm cr}$ is likely to be dominated by the stellar mass of the foreground galaxy, and hence is known from observations except for a factor 
representing the stellar mass-to-light ratio, $\Upsilon$;
and {\it (c)} the IFU allows for fairly unambiguous identification of faint counter-images, even close to the centre of the target galaxy, through the spectral contrast 
advantage of narrow emission lines.

We note here that \cite{2015ApJ...803...71S} have previously described an analysis of SLACS `grade-C' systems, i.e. those with no identifiable counter-images, with the aim of
deriving constraints from singly-imaged background sources.
Their method is based on pixel-based fitting to residual (lens-subtracted) images, assuming a 
Sersic-profile source, with the lens treated as a singular isothermal ellipse, parametrized
by the normalisation $b_{\rm SIE}$. 
This approach can be effective where the background image shows distinct curvature around the foreground galaxy, which can be unambiguously attributed to lensing
(e.g. SDSSJ0847+2925, SDSSJ1446+4943 from their figures A1--A2).
In other cases, however, the pixelized method may be susceptible to over-fitting of the intrinsic source structure, or residuals from subtracting the foreground galaxy.  
For example, in around half of their grade-C systems (e.g. SDSSJ1039+1555, SDSS0818+5410), the residual image shows a single source without 
obvious distortions suggestive of lensing, yet the Shu et al. analysis surprisingly still recovers $b_{\rm SIE}$ with a few per cent precision. 
The origin of the tight constraint, especially the lower bound, is unclear in these cases.
By comparison, our `point image' method, using only an estimated upper limit on the counter-image flux, is more conservative but probably more robust. 
In any case, the unresolved  images in our example cases are not suitable for pixel-based fitting methods.

The results of Section~\ref{sec:applic} are intended to demonstrate the potential of the upper-limit lensing method, rather than to be interpreted as firm limits on the 
IMF in massive galaxies. In particular, we have not attempted here to determine the `correct' reference mass-to-light ratio for each galaxy. 
The three examples show the variety of results which can be obtained. In SNL-4, the constraint is currently weak, and can only be tightened with 
much deeper observations, due to the faintness of the observed source and the alignment of the background galaxy with the major axis of the lens.
For PGC007748, a useful (though not very restrictive) limit on the IMF is already possible: 
the dynamical $\alpha$-versus-$\sigma$ relation of \cite{2013MNRAS.432.1862C}
predicts $\alpha$\,$=$\,1.5$\pm$0.3 at the velocity dispersion of this galaxy (254\,km\,s$^{-1}$); hence our limit of $\alpha$\,$\la$\,1.4 (if the stellar population is old) places
PGC007748 below the mean of this relation.
Finally in J0728+4005, the existing MaNGA data do not provide a useful constraint, but realistic future observations should establish a firm limit on the IMF in this object, 
since a detectable counter-image should be present over a wide range of $\Upsilon$. Conversely, failure to detect a counter-image in deeper data would imply $\alpha$\,$\la$\,1.0, 
which would be $\sim$2$\sigma$ below the Cappellari et al. trend at this velocity dispersion.

These results lead to two related applications for the upper-limit lensing approach described here.  

First, and most obviously, exploiting singly-imaged background sources promises to increase the sample of low-redshift galaxies amenable to lensing constraints on $\alpha$. In particular, 
while large IFU galaxy surveys like SAMI and MaNGA should generate a few new multiple-image lens systems
\citep{2017MNRAS.464L..46S,2018MNRAS.477..195T}, the number of singly-imaged close-projected emitters will be much larger. 
In principle, the inclusion of upper limits can provide additional information on the intrinsic distribution of $\alpha$, e.g. using survival statistics methods \citep{1985ApJ...293..192F}.
Indeed, including constraints from singly-imaged systems is {\it essential} to avoid a `lensing bias' which would otherwise favour higher $\alpha$ galaxies at a given separation.
\cite{2015ApJ...803...71S} make a similar point with regard to their derivation of the total mass profile slopes from SLACS.

\begin{figure}
\includegraphics[width=80mm,angle=270]{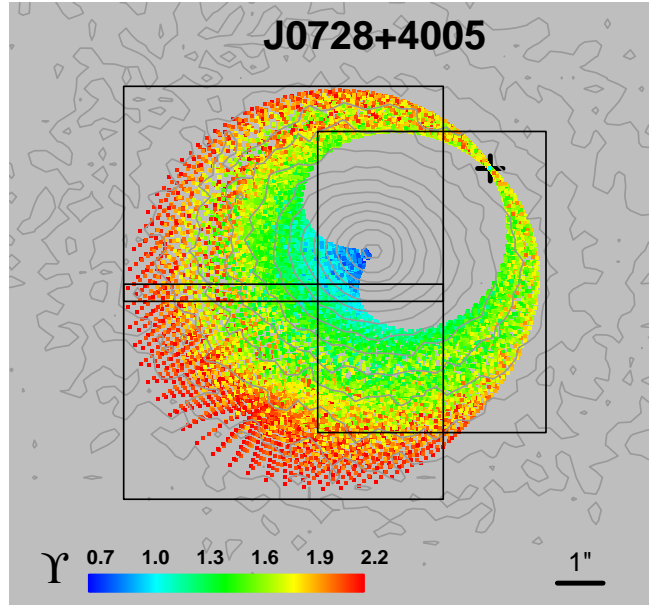}
\vskip 2mm
\caption{Predicted counter-image locations from the lensing model grid, as a function of the stellar mass-to-light ratio, $\Upsilon$,
for a deep observation of J0728+4005, with $f_{\rm lim}$\,=\,0.1 (see Figure~\ref{fig:unlens_m}).
The `test image' (the position of the detected image of the background emitter) is shown by the black cross. The 5$\times$7\,arcsec$^2$ rectangles
indicate hypothetical future Gemini/GMOS IFU observations motivated by our analysis.}
\label{fig:pred}
\end{figure}

The second application of our method is as a framework to guide follow-up observations of systems with close projected background sources, with the aim of either {\it (a)} recovering a faint
counter-image, to establish a new lens system, or {\it (b)} attaining more stringent upper-limit lensing constraints, by pushing $U(\Upsilon)$ closer to the fraction of intrinsically-single sources.
This is exemplified by the case of J0728+4005. 
Figure~\ref{fig:pred} shows the positions of the counter-images predicted from the model grid for this galaxy, colour-coded by $\Upsilon$, and the suggested
arrangement of three GMOS IFU fields,  as used in our $f_{\rm lim}$\,=\,0.1 calculation in Figure~\ref{fig:unlens_m}.

Taken together, these strategies offer a promising route to speed up the hitherto laborious task of enlarging the sample of low-redshift lenses, with the aim of deriving secure and robust 
limits on variation in the IMF mass excess factor in massive early-type galaxies. In future work, we will present a combined analysis of multiple- and single-image systems from ongoing 
observational programmes, including a more rigorous treatment of the conversion between $\Upsilon$ and $\alpha$.

\section*{Acknowledgements} 

RJS and JRL acknowledge support from the STFC through grant ST/P000541/1.
WPC was supported by STFC studentship ST/N50404X/1.
All datasets used in this paper are publicly available from the corresponding observatory archives.
This work is based in part on observations collected at the European Organisation for Astronomical Research in the Southern Hemisphere under ESO programmes 093.B-0193, 095.B-0736 and 096.B-0832.
The Pan-STARRS1 Surveys (PS1) and the PS1 public science archive have been made possible through contributions by the Institute for Astronomy, the University of Hawaii, the Pan-STARRS Project Office, the Max-Planck Society and its participating institutes, the Max Planck Institute for Astronomy, Heidelberg and the Max Planck Institute for Extraterrestrial Physics, Garching, The Johns Hopkins University, Durham University, the University of Edinburgh, the Queen's University Belfast, the Harvard-Smithsonian Center for Astrophysics, the Las Cumbres Observatory Global Telescope Network Incorporated, the National Central University of Taiwan, the Space Telescope Science Institute, the National Aeronautics and Space Administration under Grant No. NNX08AR22G issued through the Planetary Science Division of the NASA Science Mission Directorate, the National Science Foundation Grant No. AST-1238877, the University of Maryland, Eotvos Lorand University (ELTE), the Los Alamos National Laboratory, and the Gordon and Betty Moore Foundation.
This research has made use of the NASA/IPAC Extragalactic Database (NED), which is operated by the Jet Propulsion Laboratory, California Institute of Technology, under contract with the National Aeronautics and Space Administration.

Funding for the Sloan Digital Sky Survey IV has been provided by the Alfred P. Sloan Foundation, the U.S. Department of Energy Office of Science, and the Participating Institutions. SDSS-IV acknowledges
support and resources from the Center for High-Performance Computing at
the University of Utah. The SDSS web site is www.sdss.org.

SDSS-IV is managed by the Astrophysical Research Consortium for the 
Participating Institutions of the SDSS Collaboration including the 
Brazilian Participation Group, the Carnegie Institution for Science, 
Carnegie Mellon University, the Chilean Participation Group, the French Participation Group, Harvard-Smithsonian Center for Astrophysics, 
Instituto de Astrof\'isica de Canarias, The Johns Hopkins University, 
Kavli Institute for the Physics and Mathematics of the Universe (IPMU) / 
University of Tokyo, the Korean Participation Group, Lawrence Berkeley National Laboratory, 
Leibniz Institut f\"ur Astrophysik Potsdam (AIP),  
Max-Planck-Institut f\"ur Astronomie (MPIA Heidelberg), 
Max-Planck-Institut f\"ur Astrophysik (MPA Garching), 
Max-Planck-Institut f\"ur Extraterrestrische Physik (MPE), 
National Astronomical Observatories of China, New Mexico State University, 
New York University, University of Notre Dame, 
Observat\'ario Nacional / MCTI, The Ohio State University, 
Pennsylvania State University, Shanghai Astronomical Observatory, 
United Kingdom Participation Group,
Universidad Nacional Aut\'onoma de M\'exico, University of Arizona, 
University of Colorado Boulder, University of Oxford, University of Portsmouth, 
University of Utah, University of Virginia, University of Washington, University of Wisconsin, 
Vanderbilt University, and Yale University.

\bibliographystyle{mnras}
\bibliography{rjs}

\bsp	
\label{lastpage}
\end{document}